\documentclass[amsmath,amssymb,aps,prl,twocolumn,superscriptaddress]{revtex4-2}
\usepackage{graphicx}
\usepackage{bm}

\begin{document}
\title{Topological pumping in origami metamaterials}

\author{Shuaifeng Li}
\affiliation{Department of Aeronautics and Astronautics, University of Washington, Seattle, WA 98105, USA}
\affiliation{Department of Physics, University of Michigan, Ann Arbor, MI 48109, USA}
\author{Panayotis G. Kevrekidis}
\affiliation{Department of Mathematics and Statistics, University of Massachusetts, Amherst, MA 01003, USA}
\author{Xiaoming Mao}
\affiliation{Department of Physics, University of Michigan, Ann Arbor, MI 48109, USA}
\author{Jinkyu Yang}
\affiliation{Department of Aeronautics and Astronautics, University of Washington, Seattle, WA 98105, USA}
\affiliation{Department of Mechanical Engineering, Seoul National University, Seoul 08826, Republic of Korea}

\begin{abstract}
In this study, we present a mechanism of topological pumping in origami metamaterials with spatial modulation by tuning the rotation angles. Through coupling spatially modulated origami chains along an additional synthetic dimension, the pumping of waves from one topological edge state to another is achieved, where the Landau-Zener transition is demonstrated by varying the number of coupled origami chains. Besides, the inherent nonlinearity of origami metamaterials enable the excitation-dependent Landau-Zener tunneling probability. Furthermore, with the increase of nonlinearity, the topological states tend to localize in several regions in a way reminiscent of discrete breathers. Our findings pave the way towards inter-band transitions and associated topological pumping features in origami metamaterials.
\end{abstract}

\maketitle

\textit{Introduction.}--Topological metamaterials have 
had a remarkable recent impact not only in condensed matter physics but also through wave manipulation in photonics and phononics, underpinned by the robustness to imperfections~\cite{ni2023topological,ozawa2019topological,xue2022topological,ma2019topological,huang2021recent}. A recent research direction has focused on the higher dimensional topological effects in lower dimensional systems by exploiting synthetic dimensions in the parameter space~\cite{lustig2019photonic,dutt2020higher,chen2021creating,fan2019probing}. Among these pursuits, Thouless pumping stands out as a captivating demonstration of topological phenomena, offering a dynamic counterpart to the two-dimensional quantum Hall effect. Thouless pumping in photonics~\cite{verbin2015topological,kraus2012topological,ke2016topological,xu2023topological}, acoustics~\cite{ni2019observation,chen2021topological,chen2021landau}, and elasticity~\cite{apigo2018topological,xia2020topological,rosa2019edge,riva2020edge,riva2020adiabatic,wang2023smart} not only  illuminates the topological aspects of dynamic evolution but also highlights its significance in unraveling the higher-dimensional topological physics; see also the recent review~\cite{citro2023thouless}.

Meanwhile, recently the ancient art of origami, with its intricate paper-folding techniques, has captured the interest of the physics and engineering communities~\cite{meloni2021engineering,peraza2014origami,zhu2022review}. Origami metamaterials, comprising an assemblage of origami blocks, offer a unique repertoire of static properties and the dynamic characteristics that hold the promise of mitigating impacts and controlling vibrations~\cite{lyu2021origami,ji2021vibration,yasuda2019origami,zhou2016dynamic}. Notably, origami metamaterials have been proven to demonstrate the potential for the realization of systems of relevance to condensed matter physics, particularly in the exploration of topological states~\cite{miyazawa2022topological,li2023geometry,li2023elastic}. Origami metamaterials, coupled with their distinct mechanical properties being tuned through initial configurations, position them as convenient platforms for the realization of Thouless pumping, thus opening up intriguing possibilities. Additionally, while most studies on band topology are conducted in linear systems, there is growing interest in the investigation of properties in the presence of nonlinearities~\cite{smirnova2020nonlinear,maczewsky2020nonlinearity,darabi2019tunable,ma2023nonlinear}. Therefore, the intrinsic nonlinearity in origami offers opportunities for investigating the interplay between nonlinearity and topological pumping in elastic systems, which warrants further study.

In this work, we demonstrate topological Thouless pumping in an origami metamaterial consisting of coupled Kresling origami chains, each tailored with modulated geometrical parameters. By engineering the initial rotation angles of the origami units according to the Aubry-Andr\'e-Harper~(AAH) model~\cite{harper1955single,harper1955general}, we find nontrivial topological edge states in a single origami chain. Leveraging an additional parameter as a synthetic dimension, we realize the Landau-Zener transition~\cite{landau,zener} and topological pumping in the two-dimensional system with coupled modulated origami chains. Furthermore, our exploration reveals that the nonlinearity of origami can result in asymmetrical Landau-Zener tunneling in a way reminiscent of cold atom experiments~\cite{jona2003asymmetric} (but also with some differences discussed herein), and also lead to the transition from topological pumping to discrete breathers~\cite{flach1998discrete} and bulk states. Our work not only achieves topological pumping in origami metamaterials but also delves into the profound influence of nonlinearity thereon.

\textit{Design of origami metamaterials.}--We first consider the single Kresling origami chain that is designed following the AAH model. The on-site potentials of AAH model, specifically for the origami chain, stiffness values are spatially modulated by the initial rotation angle of Kresling origami $\theta_{x_{n}}=\theta_{0}+\mu\theta_{0}\sin(2\pi x_{n}\xi+\eta)$. Here, $\mu$, $\xi$ and $\eta$ represent the modulation amplitude, modulation frequency and modulation phason~(also the synthetic dimension of the system). In our work, we choose $25$ Kresling origami~($\{x_{n}\in\mathbb{Z}|0\leq x_{n}\leq 24\}$), $\theta_{0}=70^{\circ}$ and $\mu=0.2$. Thus, there are $n=26$ separators, each of which has two degrees of freedom~(translation $u$ and rotation $\varphi$) according to the truss model~\cite{yasuda2019origami,miyazawa2022topological}, where massless trusses and no bending of the chain are considered. More details about geometry, parameters and truss model are shown in the Supplemental Material~\cite{supplementary}.
\begin{figure*}[ht]
    \centering
    \includegraphics[width=0.8\textwidth]{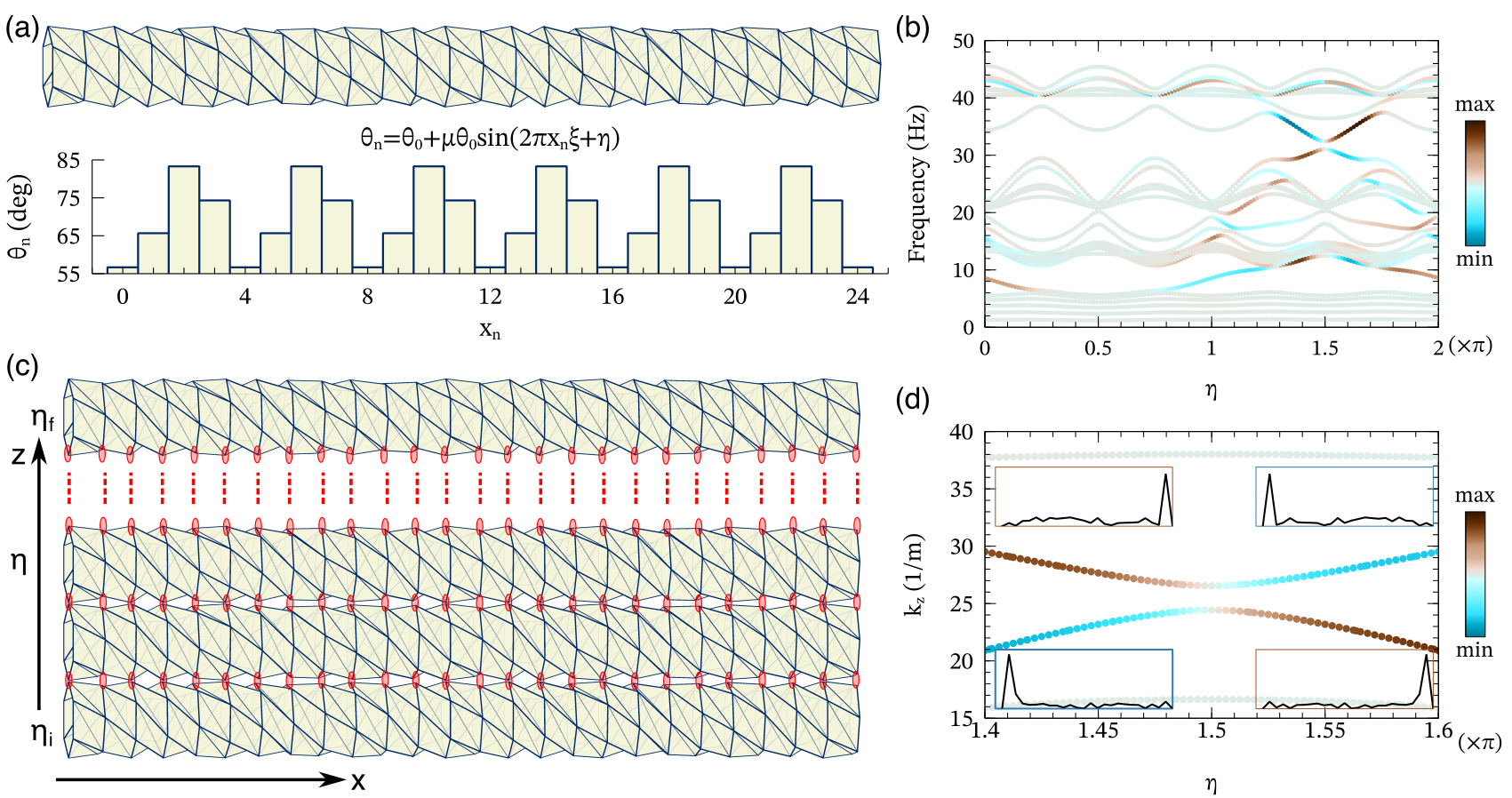}
    \caption{(a) The side view of the the Kresling origamis chain with spatial modulation. The rotation angle of each Kresling origami follows $\theta_{x_{n}}=\theta_{0}+\mu\theta_{0}\sin{(2\pi x_{n}\xi+\eta)}$. The corresponding values are shown below.
    (b) The calculated eigenspectrum as a function of $\eta$.
    (c) The coupled origami chains via torsional springs $k_{c}$ along a synthetic dimension $\eta$. $\eta$ varies from $\eta_{i}$ to $\eta_{f}$. The red markers represent the torsional springs.
    (d) The wave number $k_{z}$ as a function of $\eta$ for a fixed frequency $f=40~\mathrm{Hz}$. The insets show four eigenmodes~($\varphi$) of the origami chain along $x$ direction, corresponding to $\eta_i=1.4\pi$ and $\eta_f=1.6\pi$ of two colored bands, respectively.
    The color in (c) and (d) represents the localization index, indicating the localization of the energy.
    \label{fig:fig1}
    }
\end{figure*}

One of the properties about the AAH model is the
fractal nature of its spectrum. To explore this, we show the spectrum as a function of the parameter $\xi$ and observe a structure known as Hofstadter butterfly~\cite{hofstadter1976energy}.The Hofstadter butterfly is not only a footprint of the spectral fractality, but also the indicator of a nontrivial bandgap whose gap Chern number is conveniently calculated by the integrated density of states~(IDS)~\cite{prodan2019k}~(see Supplemental Material~\cite{supplementary}). In our single Kresling origami chain displayed in FIG.~\ref{fig:fig1}, we choose $\xi=\frac{1}{4}$ to study the topological states in the nontrivial bandgap with the gap Chern number $C_{g}=-1$, and the rotation angle of each origami in the chain is exhibited in bottom panel of FIG.~\ref{fig:fig1}(a). In this way, this one-dimensional system satisfies the commensurate AAH model and each unit cell contains $4$ origamis. The Kresling origami has the typical nonlinear behaviors that the stiffness changes under different levels of deformation~(see the governing equation and force-displacement curves in Supplemental Material~\cite{supplementary}), which will be detailed later to show its effect to topological pumping. Therefore, to reveal the dispersion topology in such origami chain, we linearize the governing equation of Kresling origami to obtain linear coefficients~(Supplemental Material~\cite{supplementary}). In FIG.~\ref{fig:fig1}(b), the eigenspectrum is plotted as a function of $\eta$ with the clamped boundary condition on both sides of the origami chain. To check if the eigenmodes own features of topological states, the localization index~(LI), which is the product of the inverse participation ratio~(IPR) and the center of mode~(CoM), is used to characterize the localization of energy~\cite{miyazawa2022topological}:
\begin{eqnarray}
    \label{equ:1}
    LI=IPR\times CoM=|(\bm{K}\bm{U}^{\circ 2})^{\circ 2}|(\frac{2}{n-2}\bm{w}^{T}\bm{K}\bm{U}^{\circ 2})
\end{eqnarray}
where $\bm{U}$ is the eigenvector containing translation $u$ and rotation $\varphi$, $\bm{K}$ is the commutation matrix and $w$ is the weighting vector. $\circ$ denotes the Hadamard product. The clamped boundary condition leads to $n-2$ effective separators. The introduction of LI indicates that if the eigenmode is skewed toward the left~(right) boundary, LI is negative~(positive). More details including the definition of $\bm{K}$ and $w$ can be found in the Supplemental Material~\cite{supplementary}. According to the color-encoded spectrum in FIG.~\ref{fig:fig1}(b), topological edge states with the energy localized on either boundary can be found in the bandgap when $\eta$ is in certain range.

We choose the topological states within the bandgap around $30~\mathrm{Hz}$ when $\eta$ is between $1.4\pi$ and $1.6\pi$ to further explore. The parameter $\eta$ is used as a synthetic dimension to construct the two-dimensional origami metamaterials by coupling spatially modulated origami chains~($z_{n}$ is used to denote the number of chains) along the $z$ direction using torsional springs $k_{c}$ which couple rotations $\varphi$ only, as illustrated by FIG.~\ref{fig:fig1}(c). $\eta$ linearly varies from $\eta_{i}=1.4\pi$ to $\eta_{f}=1.6\pi$. Variation of wavenumber $k_{z}$ shifts the entire spectrum~[FIG.~\ref{fig:fig1}(b)] by considering slowly-varying $\eta$ along the $z$ direction, resulting in the spectrum $f(\eta,k_{z})$. We then fix the excitation frequency $f(\eta,k_{z})=40~\mathrm{Hz}$ in the following discussion and generate the relation between $k_{z}$ and $\eta$ shown in FIG.~\ref{fig:fig1}(d). Topological edge states are illustrated with large $|LI|$~(colored), while the bulk bands are illustrated with small $|LI|$~(gray). Moreover, the eigenmodes~($\varphi$) in the insets clearly exhibit the topological edge states with localized energy on either left end or right end.

\begin{figure}[ht]
    \centering
    \includegraphics[width=0.48\textwidth]{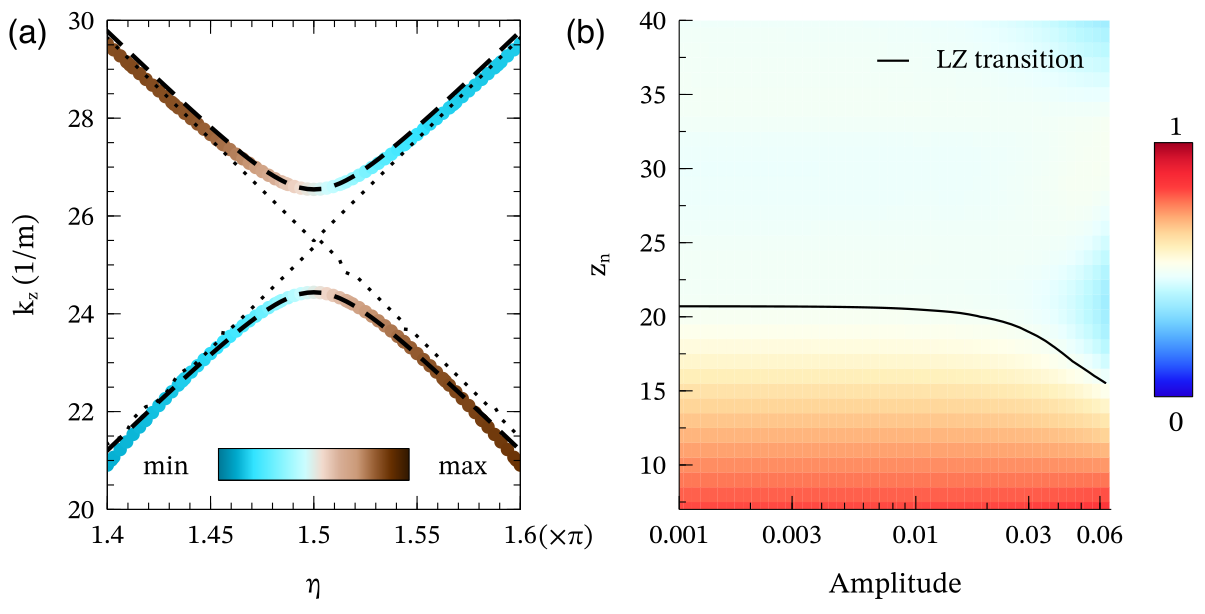}
    \caption{(a) The close-up view of the relation between $k_{z}$ and $\eta$. The black dashed lines and black dotted lines represent the theoretical adiabatic evolution and diabatic evolution, respectively.
    (b) The Landau-Zener tunneling probability as a function of number of chains and excitation amplitude. The black solid lines indicate the Landau-Zener transition point, $P_{|\Psi_{L}\rangle}=0.5$.
    \label{fig:fig2}
    }
\end{figure}

\textit{Landau-Zener tunnelling in origami metamaterials.}--Of particular interest are two bands featuring topological edge states. As depicted in the close-up view in FIG.~\ref{fig:fig2}(a), there is a bandgap with a size of $\Delta k_{z}=2.1~\mathrm{m^{-1}}$. This small size of the bandgap suggests the sensitivity of topological pumping to the condition of adiabaticity and hence offers a unique opportunity to explore diabatic transition. In the vicinity of $\eta=1.5\pi$, these two bands can be modeled by a two-level effective Hamiltonian near $k_{z}=25.5~\mathrm{m^{-1}}$ by considering the $z$ direction as time:
\begin{eqnarray}
    H(\delta\eta)=\begin{pmatrix}
        -\alpha\delta\eta & \Delta k_{z}/2 \\
        \Delta k_{z}/2 & \alpha\delta\eta
    \end{pmatrix}
\end{eqnarray}
where $\alpha=12.9~\mathrm{m^{-1}}$ serving as a fitting parameter between theoretical model and simulation by minimizing the mean squared error. The bases of $H$ are denoted as $|\Psi_{L}\rangle$ and $|\Psi_{R}\rangle$, representing the topological edge states localized at the left and right boundaries, respectively. The eigenvalues of $H$ are illustrated as the black dashed lines in FIG.~\ref{fig:fig2}(a), showcasing the excellent agreement with the results from simulation~(colored dots). If the initial state is $|\Psi_{L}\rangle$ at $\eta=1.4\pi$, residing in the lower level, the final state $|\Psi_{f}\rangle$ will be a superposition of two topological edge states $|\Psi_{L}\rangle$ and $|\Psi_{R}\rangle$. Following the evolution of $|\Psi_{L}\rangle$, the final state remains localized on the left boundary, corresponding to the diabatic transition~(black dotted lines). Alternatively, following the lower level during pumping, the final state will be dominated by $|\Psi_{R}\rangle$, leading to the localization on the right boundary.

The composition of the final state $|\Psi_{f}\rangle$ can be predicted by the Landau-Zener model~\cite{shevchenko2010landau}, given by $|\Psi_{f}\rangle=L(z)|\Psi_{L}\rangle+R(z)|\Psi_{R}\rangle$, with $L(z)$ and $R(z)$ satisfying the following relation:
\begin{eqnarray}
    i\frac{d}{dz}\begin{pmatrix}
        L(z)\\R(z)
    \end{pmatrix}=\begin{pmatrix}
        -\beta z & \Delta k_{z}/2\\
        \Delta k_{z}/2 & \beta z
    \end{pmatrix}
    \begin{pmatrix}
        L(z)\\R(z)
    \end{pmatrix}
\end{eqnarray}
Here, $\beta=\alpha(\Delta\eta/z_{n})$ represents the adiabaticity, which depends on the rate of parameter evolution, $\Delta\eta/z_{n}$, with $\Delta\eta=0.2\pi$ being the span of $\eta$. By assuming the initial state to be $|\Psi_{L}\rangle$, the proportion of final state can be derived as: $P_{|\Psi_{L}\rangle}=L^{2}(z_{n})=e^{-\pi \Delta k_{z}^{2}/4\beta}$ and $P_{|\Psi_{R}\rangle}=R^{2}(z_{n})=1-e^{-\pi \Delta k_{z}^{2}/4\beta}$. In Supplemental Material~\cite{supplementary}, we show $P_{|\Psi_{L}\rangle}$ as a function of $z_{n}$. Therein, when $P_{|\Psi_{L}\rangle}=0.5$, the Landau-Zener transition point corresponds to $z_{n}=21$. When $z_{n}$ is sufficiently large, the evolution is slow enough to be adiabatic, resulting in the final state dominated by $|\Psi_{R}\rangle$. However, when the same process occurs in a system with a smaller $z_{n}$, the variation is fast, leading to diabatic behaviors and the final state dominated by $|\Psi_{L}\rangle$.

As mentioned in the introduction, Kresling origami exhibits intrinsic nonlinearity. Therefore, we investigate the Landau-Zener tunneling probability calculated by the kinetic energy of the separator as a function of the excitation amplitude. As shown in FIG.~\ref{fig:fig2}(b), when the excitation amplitude is small and the system remains close to the linear regime with the initial state $|\Psi_{L}\rangle$ in the lower band, the Landau-Zener transition point~($P_{|\Psi_{L}\rangle}=0.5$, black solid line) remains consistent~($z_{n}=21$), showing agreement with theoretical results. However, as the amplitude increases to the weakly nonlinear regime, the transition point shifts to the smaller $z_{n}$. This suggests that the tunnelling probability decreases when the excitation amplitude increases. While previous studies have shown that in the presence of nonlinearity the tunnelling probability will increase due to the interactions between the particles~\cite{liu2002theory,jona2003asymmetric,konotop2005landau}, our results show the opposite case, which may result from the strain softening behaviors of Kresling origami. Note that the discussion is only limited in the weakly nonlinear regime because larger excitation amplitude beyond the limit as presented will eliminate topological states, which will be shown below.

\textit{Transition of topological pumping.}--To further explore the nonlinear effects of origami on the topological pumping, we excite the system with varying amplitudes from the left boundary to obtain the initial state $|\Psi_{L}\rangle$ in the lower band. We also fix $z_{n}=101$ to ensure the adiabatic evolution, resulting in the final state being predominantly composed of $|\Psi_{R}\rangle$.

\begin{figure}[ht]
    \centering   
    \includegraphics[width=0.48\textwidth]{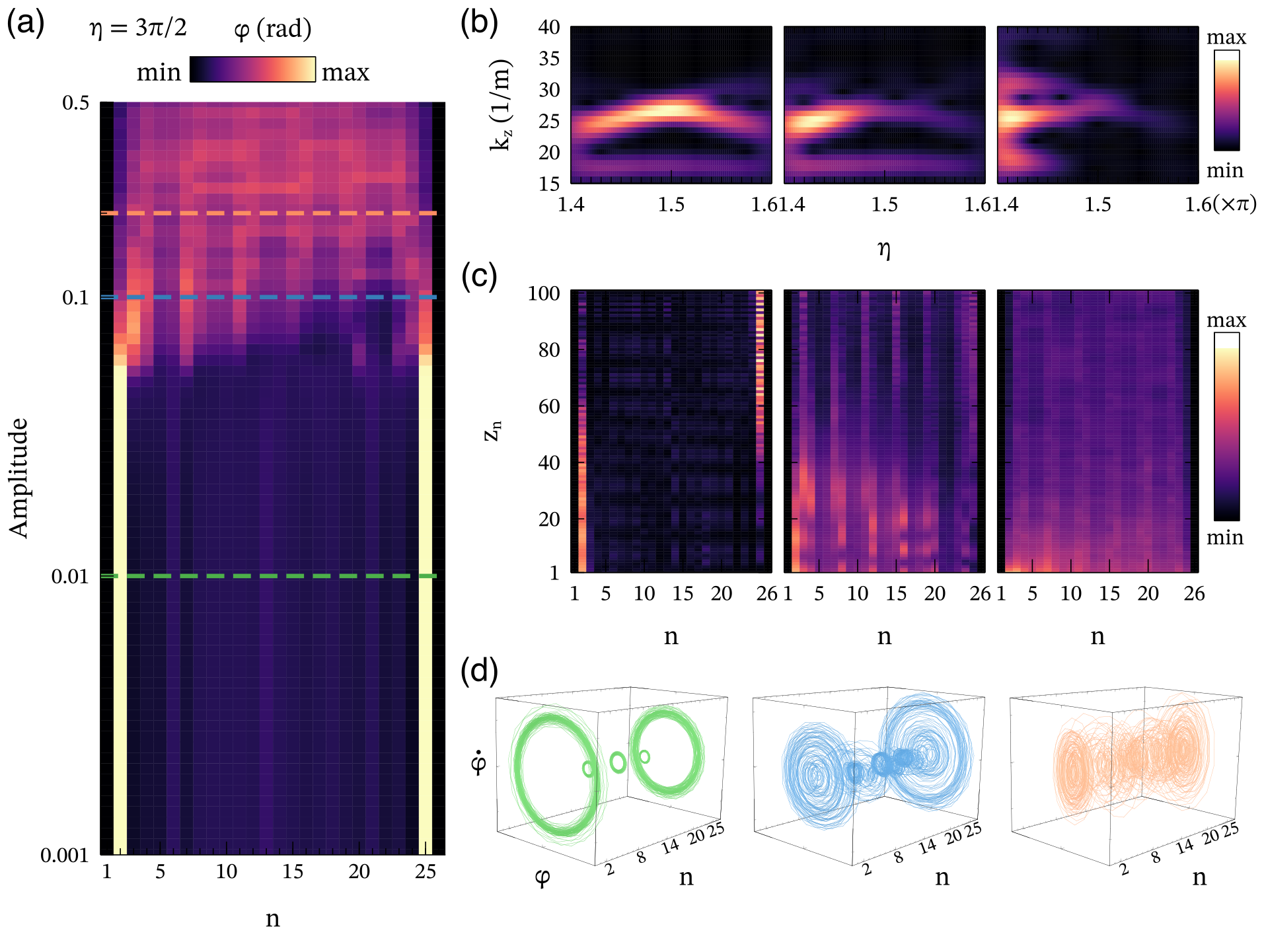}
    \caption{(a) The RMS of $\varphi$ when $\eta=3\pi/2$ as a function of excitation amplitude.
    (b) The spectrogram of the time domain simulation, showcasing the relation between $\eta$ and $k_{z}$.
    (c) The RMS of $\varphi$ under different excitation amplitudes.
    (d) Phase space representation plot~($\varphi$ vs $\dot{\varphi}$) at different sites when $\eta=3\pi/2$.
    From left to right, the panels in (b), (c) and (d) indicates the cases with increasing excitation amplitudes, corresponding to the dashed lines in (a).
    \label{fig:fig3}}
\end{figure}

In FIG.~\ref{fig:fig3}(a), we show the the root mean square~(RMS) of field distribution~($\varphi$) of the $50$th origami chain~(middle chain), where $\eta=3\pi/2$, as a function of excitation amplitude. Evidently, when the amplitude is small~(near the linear regime), topological states can be successfully transferred from the left to the right boundary with the minimal involvement of the bulk states. To illustrate the adiabatic nature of the pump, we employ a time-frequency analysis~(spectrogram) to represent the rotational displacement field in the reciprocal space $\hat{\varphi}(\eta,k_{z},k_{x},f)$. For visualization purposes, $f=40~\mathrm{Hz}$ and RMS along the $k_{x}$ dimensions are taken, which produces $\hat{\varphi}(\eta,k_{z})$. As shown in the first panel of FIG.~\ref{fig:fig3}(b), when the amplitude is $0.01$, the wave is initially distributed around wave numbers $k_{z}$ corresponding to the excited left-localized topological state. Along the synthetic dimension $\eta$, it closely follows the evolution of the lower band. At the end of the process, the majority of the energy is concentrated on the right-localized mode, with a portion scattered to neighboring bulk states. The corresponding RMS of $\varphi$ is shown in the first panel of FIG.~\ref{fig:fig3}(c), providing a clear depiction of the topological pumping. In addition, the phase space representation~[$\varphi(t)$ vs $\dot{\varphi}(t)$] is displayed in the first panel of FIG.~\ref{fig:fig3}(d) for several separators along the $50$th origami chain, showcasing periodic orbits in the phase space. At small amplitude, these mainly involve such orbits at the two boundaries.

As the excitation amplitude continues to rise, the influence of nonlinearity becomes prominent. Gradually, the dominance of topological pumping wanes, bypassing the boundary-induced confinement and exploring more widely the configuration space of the system, as shown in FIG.~\ref{fig:fig3}(a). As an example indicated by blue dashed line, when the amplitude reaches $0.1$, $\hat{\varphi}(\eta,k_{z})$ is shown in the second panel of FIG.~\ref{fig:fig3}(b). In comparison to the linear case, it becomes evident that the energy is not primarily concentrated on the right-localized mode by the end of the evolution. The RMS of $\varphi$ is displayed in the second panel of FIG.~\ref{fig:fig3}(c), revealing the emergence of localized states reminiscent of discrete breathers~\cite{flach1998discrete}, principally located at certain sites~($3$rd, $7$th, $\cdots$, $23$rd separators). These modes have certain frequencies located at the bandgap~(in line with what is expected based on their spatial localization). Further analysis of the sub-harmonic effect can be found in Supplemental Material~\cite{supplementary}. Likewise, the phase space representation is illustrated in the second panel of FIG.~\ref{fig:fig3}(d). Compared with the linear case, the trajectories show nonlinear features on the periodic orbits and the corresponding excitation of larger amplitude orbits in the bulk of the system. We note that the similar transitions from topological states to discrete breathers has also been observed in the one-dimensional spatially modulated nonlinear spring chain~\cite{rosa2022amplitude}.

If the excitation amplitude is increased further, the influence of nonlinearity becomes increasingly pronounced. The bulk states will be excited, as evidenced by the large amplitude of $\varphi$ in the bulk shown in FIG.~\ref{fig:fig3}(a) when the amplitude is $0.2$ marked by the orange dashed line. The third panel of FIG.~\ref{fig:fig3}(b) and additional details in the Supplemental Material~\cite{supplementary} further confirm that a significant portion of the energy is scattered to the bulk, at $40~\mathrm{Hz}$ and other frequencies. According to the RMS of $\varphi$ in the third panel of FIG.~\ref{fig:fig3}(c), the characteristic feature of discrete breathers, where energy is localized at specific sites, becomes less apparent. Instead, the bulk of the system experiences a progressively more uniform excitation. In stark contrast to the previous cases, the relation between $\varphi$ and $\dot{\varphi}$ exhibits highly nonlinear features, exploring the full phase 
space and significantly deviating from well-defined periodic orbits, as displayed in the third panel of FIG.~\ref{fig:fig3}(d).

In conclusion, our study demonstrates the design of origami metamaterials by coupling spatially modulated Kresling origami chains, following the AAH model, to achieve topological pumping. The intrinsic nonlinearity of origami metamaterials reveals an intriguing phenomenon, characterized by Landau-Zener tunneling probability contingent upon the excitation amplitude~(in a way reflecting
the system's strain softening nonlinearity). Moreover, as the nonlinearity becomes more pronounced, the topological states gradually transform into spatially localized, temporally periodic states reminiscent of discrete breathers and eventually transition into progressively more uniform bulk states. Our findings provide valuable insights into the evolutionary path from topological pumping to localized, breathing and bulk states in nonlinear systems. These findings provide insights into potential applications in origami-based architectures for the manipulation of elastic waves since parameters used in simulations are from previous experimental work~\cite{miyazawa2022topological}, and the exploration of Landau-Zener-St{\"u}ckelberg interferometry as well as the corresponding nonlinear features~\cite{shevchenko2010landau,li2018nonlinear}.

\begin{acknowledgments}
\noindent \textit{Acknowledgments}.--S.L. and J.Y. are grateful for the support from the U.S. National Science Foundation~(Grant No. 1933729 and 2201612). S.L. and X.M. thank the support from the Office of Naval Research~(MURI N00014-20-1-2479). This material is based upon work supported by the U.S.\ National Science Foundation under the awards PHY-2110030 and DMS-2204702 (P.G.K.). J.Y. acknowledges the support from SNU-IAMD, SNU-IOER, and National Research Foundation grants funded by the Korea government: 2023R1A2C2003705 and 2022H1D3A2A03096579~(Brain Pool Plus by the Ministry of Science and ICT).

\end{acknowledgments}

\bibliographystyle{apsrev4-2}
\bibliography{references}

\end{document}